\documentclass{emulateapj}
\usepackage{apjfonts}
\slugcomment{Accepted by  \textsc{the Astrophysical Journal}}

\newcommand{\tauc}{\tau_{\mathrm{c}}}

\newcommand{\ppt}[1]{\frac{\partial #1}{\partial t}}
\newcommand{\ddt}[1]{\frac{d #1}{d t}}
\newcommand{\bfv}{{\bf v}}
\newcommand{\bffr}{{\bf F_{\mathrm{R}}}}
\newcommand{\fr}{{F_{\mathrm{R}}}}
\newcommand{\fc}{{F_{\mathrm{C}}}}
\newcommand{\frbar}{{\bar{F}_{\mathrm{R}}}}
\newcommand{\fcbar}{{\bar{F}_{\mathrm{C}}}}
\newcommand{\bffrbar}{{\bf \bar{F}_{\mathrm{R}}}}
\newcommand{\bffcbar}{{\bf \bar{F}_{\mathrm{C}}}}
\newcommand{\rhobar}{\bar{\rho}}
\newcommand{\delrho}{\Delta \rho}
\newcommand{\pbar}{\bar{p}}

\newcommand{\delp}{\Delta p}
\newcommand{\vbar}{{\bf u}}
\newcommand{\delv}{{\bf V}}
\newcommand{\Tbar}{\bar{T}}
\newcommand{\delT}{\Delta T}
\newcommand{\xhat}{\hat{{\bf x}}}
\newcommand{\yhat}{\hat{{\bf y}}}
\newcommand{\zhat}{\hat{{\bf z}}}
\newcommand{\sbar}{\bar{s}}
\newcommand{\dels}{\Delta s}
\newcommand{\bfk}{{\bf k}}
\newcommand{\barh}{\bar{h}}
\newcommand{\kB}{k_{\mathrm{B}}}

\newcommand{\fconv}{f_{\mathrm{c}}}
\newcommand{\dtovert}{\frac{\delta \Tbar}{\Tbar}}
\newcommand{\dpoverp}{\frac{\delta \pbar}{\pbar}}
\newcommand{\drhooverrho}{\frac{\delta \rhobar}{\rhobar}}
\newcommand{\dsovers}{\frac{\delta \sbar}{\sbar}}

\begin{document}

\title{Damping of Type I X-ray Burst Oscillations by Convection}
\shorttitle{Convection and Burst Oscillations}

\author{Randall L.\ Cooper}
\shortauthors{Cooper}
\affil{Harvard-Smithsonian Center for Astrophysics, 60 Garden 
Street, Cambridge, MA 02138}

\email{rcooper@cfa.harvard.edu}

\begin{abstract}
I construct a simple model of the convective burning layer
during a type I X-ray burst to investigate the effects convection has
on the stability of the layer to nonradial oscillations.  A
linear perturbation analysis demonstrates that the region is stable to
nonradial oscillations when energy transport is convection-dominated,
but it is unstable when energy transport is radiation-dominated.
Thus, efficient convection always dampens oscillations.  These results
may explain the nondetection of oscillations during the peak of
some X-ray bursts.
\end{abstract} 

\keywords{accretion, accretion disks --- stars: neutron --- X-rays:
binaries --- X-rays: bursts}

\section{Introduction}\label{introduction}

Type I X-ray bursts are thermonuclear explosions that occur on the
surfaces of accreting neutron stars.  Fast rises and exponential-like
decays lasting $\sim 1$ and $\sim 10$--$100$ seconds, respectively,
characterize burst lightcurves \citep[for reviews,
see][]{LvPT95,SB06}.  \citet{Setal96} detected coherent oscillations
in the lightcurve of a type I X-ray burst from the low-mass X-ray
binary 4U 1728--34.  Since then, astronomers have detected
oscillations in burst lightcurves from 19 additional sources
\citep[although some detections are unconfirmed; see][and references
therein]{B07,MKSW07,SMK08,LB08}.  Bursts can exhibit oscillations
during the rise, the peak, and/or the decay of their lightcurves.  The
oscillation frequency typically increases by a few Hz during a burst,
but it asymptotes to a specific frequency unique to that source to
within a few parts in $10^{3}$ \citep[e.g.][]{GHSC02,MCGP02}.  The
asymptotic frequency's stability implies that it corresponds to the
neutron star spin frequency \citep[e.g.][]{SM99}.  The discoveries
that the coherent oscillation frequency during a superburst from 4U
1636--536 matches the asymptotic frequency of burst oscillations from
that source \citep{SM02}, the burst oscillation frequencies from the
accretion-powered millisecond pulsars SAX J1808.4--3658
\citep{CMMGWvdKM03} and XTE J1814--338 \citep{MSS03,SMSintZ03} match
their respective known spin frequencies, and the frequency of coherent
pulsations \citet{CAPWvdK08} detected in the persistent emission from
Aql X-1 is close to the asymptotic frequency of burst oscillations
from that source confirm this implication.

It is most likely the rotational modulation of a growing hot spot on
the neutron star surface that generates oscillations during the burst
rise \citep{SZS97,SZSWL98}.  The time needed to accrete enough fuel to
trigger a thermal instability (a few hours to days) greatly exceeds
that needed to burn the fuel ($\sim$ seconds), which makes
simultaneous ignition over the entire stellar surface highly unlikely.
Ignition probably occurs at a point \citep{J78,S82}, and the resulting
hot spot grows and engulfs the stellar surface in $\sim 1$ s, the
burst rise time \citep{FW82,B95,Zetal01,SLU02}.  The
latitude-dependent Coriolis force regulates the thermonuclear flame
propagation speed and thereby the time evolution of the hot spot
\citep{SLU02,BS07}.  The accreted fuel usually ignites at the equator,
and the resulting thermonuclear flame quickly spreads in longitude and
generates an axisymmetric belt around the neutron star.  No
oscillations occur because there is no azimuthal asymmetry
\citep{SLU02}.  Ignition sometimes occurs off the equator
\citep{CN07,MW08}; the thermonuclear flame propagates in longitude
much more slowly, generating a long-lived non-axisymmetric hot spot
and hence oscillations.

The rotationally modulated hot spot model fails to explain
oscillations during the burst decay, however, since the flame has
engulfed the neutron star surface by this time, and any remnant
nonaxisymmetry from ignition would have dissipated since the cooling
time at the burst peak is much less than the burst duration.  It is
thought that excited surface modes generate nonradial oscillations in
the burst tail \citep{MT87,CB00,L04,H04,H05,LS05,PB05,PB06}.  In
particular, \citet{H04} suggests buoyant $r$-modes as the most
promising candidate.  $r$-Modes travel backward in the corotating
frame, so the observed oscillation frequency is less than the neutron
star spin frequency.  The $r$-mode frequency decreases as the surface
cools during the burst decay, which explains the observed increase in
the oscillation frequency \citep[although Heyl's model may not
adequately predict the observed frequency drift and the asymptotic
frequency's stability;][]{PB05,BL08}.  $r$-Modes occupy only a small
area around the equator, so their amplitudes much be huge to produce
the observed $\sim10\%$ flux variations.  Therefore, the modes must be
driven unstable.  \citet{NC07} propose that the $\epsilon$-mechanism
drives surface modes during the burst decay: if heating via nuclear
burning is sufficiently strong relative to cooling via radiative
diffusion and emission, some nuclear energy converts to mechanical
energy and drives the oscillations \citep[see
also][]{LB82,MT87,SL96,PB04}.  They predict that short, powerful
bursts preferentially exhibit oscillations, which is generally in
accord with observations.

If the $\epsilon$-mechanism does indeed drive oscillations, then one
would naively predict that oscillations occur preferentially when
nuclear burning is strongest, i.e. near the lightcurve's peak.  This
in not true; bursts exhibiting oscillations show them during the peak
only about half the time \citep{Getal08}.  Some bursts are so powerful
that the nuclear luminosity temporarily exceeds the local Eddington
luminosity; the extra thermal energy converts to both kinetic energy
and gravitational potential energy and expands the outermost layers of
the neutron star.  Astronomers often detect oscillations from these
so-called photospheric radius expansion (PRE) bursts
\citep{MFMB00,MCGS01}, but they rarely detect oscillations during the
PRE phase itself \citep{SMB97,SJGL97,MCGP02,Getal08}.  Why, then, do
bursts often fail to exhibit oscillations during the peak phase?

To address this question, one must consider the mechanisms by which
the burning layer transports energy.  Highly efficient convection is
the primary energy transport mechanism during the first 
few seconds of the initial phase of a
powerful burst
\citep[e.g.][]{J77,J78,HS82,WWW82,Pre83,Wetal04,FST08}. It is well
known that convection can either drive or dampen pulsations in
variable stars \citep[e.g.][and references
therein]{C80,S86,UOASS89,DGGGS05}.  Does convection affect burst
oscillations as well?  \citet{MT87} were the first to suggest such a
coupling between convection and pulsations during type I X-ray bursts,
but they did not pursue this idea.  In this investigation, I develop
an analytical one-zone model of the convective burning region during a
type I X-ray burst to determine the effects convection has on the
pulsational stability of the burning layer.  I find that efficient
convection dampens oscillations, much like it does in stars near the
red edge of the instability strip.  These results may explain the
nondetection of oscillations near the peaks of some type I X-ray
bursts.  I begin in \S \ref{oscconeq} by deriving the equations that
govern both the oscillatory and convective motions.  I perturb the
governing equations and derive the stability criterion in \S
\ref{perturbeqns}, and I apply this criterion to type I X-ray bursts
in \S \ref{pulsationalstab}.  I conclude in \S \ref{discuss} with a
discussion of the results.

\section{Oscillation and Convection Equations}\label{oscconeq}

Starting from the work of \citet{U67} and \citet{GSNB75}, I construct 
a one-zone model of the convective burning layer during a type I X-ray 
burst on a rapidly rotating neutron star.  My goal is to 
conduct a linear stability analysis on the entropy equation in 
a manner similar to that in \citet{NC07} and determine the 
stability of the configuration to nonradial oscillations.  Usually, 
one conducts such calculations assuming the unperturbed 
configuration is static.  This assumption is inapplicable to the 
problem at hand, for even the unperturbed convection zone 
involves vertical motions of the matter.  Thus it is necessary 
to distinguish between fluid motions due to convection and fluid 
motions due to oscillations.  A further complication 
is that an entirely satisfactory theory of convection does not 
yet exist.  I use the standard mixing length theory to describe 
convective motions.  Specifically, I demand that the mixing length theory of 
convection applies to both the initial configuration and the 
perturbation.  Constructing such a model is not trivial 
\citep[see, e.g.][]{GSNB75}.  Luckily, the parameters relevant to 
type I X-ray bursts allow me to simplify the problem enormously.  

I make the following assumptions: 

\noindent
(I) The Boussinesq approximation of the mixing length theory applies 
to the convective motions \citep[see, e.g.][]{CG68,HKT04}.  In particular, 
I ignore pressure fluctuations in the entropy equation and 
density fluctuations in the continuity equation.  

\noindent
(II) The mixing length $l$ is much smaller than the typical 
nonradial oscillation wavelength.  This assumption is easily 
justified, for the mixing length is presumed smaller than the 
vertical pressure scale height ($\sim$ meters), whereas the 
oscillation wavelength is of order the neutron star radius
($\sim 10$ km).

\noindent
(III) The lifetime of a convective element $\tauc$ is the 
shortest relevant timescale other than the vertical sound crossing 
timescale.  Specifically, $\tauc \ll \Omega^{-1}$, $\omega^{-1}$, 
$\tau_{\mathrm{nuc}}$, and $\tau_{\mathrm{R}}$, where $\Omega$ is the 
neutron star spin frequency, $\omega$ is the mode angular frequency, 
$\tau_{\mathrm{nuc}}$ is the nuclear burning timescale of matter 
within a convective element, and $\tau_{\mathrm{R}}$ is the 
radiative cooling timescale of a convective element.  The last 
inequality implies that convection is efficient.
Estimates of $\tauc$ range from $10^{-6}$--$10^{-3}$ s 
\citep{MT87,CB00,WBS06,FST08}, in accord with this assumption.

\noindent
(IV) Both the turbulent pressure and the rate of turbulent kinetic 
energy dissipation into heat are negligible.

\noindent
(V) The layer is always in vertical hydrostatic equilibrium.

I proceed as follows.  First, I write down the hydrodynamic equations.
These equations describe the aggregate behavior of both the
oscillatory and convective motions.  The convective motions have short
wavelengths of order the mixing length $l$, which is much smaller than
the oscillation wavelengths by assumption (II).  The disparity of the
two lengthscales allows me to separate the equations governing the
oscillatory motions from those governing the convective motions.  In this
work, I investigate only the stability of the oscillations to
nonadiabatic perturbations, not the properties of the oscillations
themselves \citep[for a discussion of the latter, see,
e.g.][]{L68,BUC96,NC07}.  Therefore, I conduct a linear stability
analysis on only the entropy equation governing the oscillatory
motion.  The perturbed entropy equation contains perturbations of
quantities describing the convective motions.  I solve for the
perturbed convective quantities in terms of quantities describing the
oscillatory motions by perturbing the convective equations.

\subsection{Fundamental Hydrodynamic Equations}

The fundamental continuity, momentum, and entropy equations of 
the model are
\begin{eqnarray}
\ppt{\rho} + \nabla \cdot (\rho \bfv) &=& 0,\label{continuityeq} \\
\ppt{\bfv} + (\bfv \cdot \nabla) \bfv + 2 \Omega \times \bfv &=& -g \zhat - \frac{1}{\rho}\nabla p,\label{momentumeq}\\
\rho T \left (\ppt{s} + \bfv \cdot \nabla s\right) &=& \rho \epsilon - 
\nabla \cdot \bffr,\label{entropyeq}
\end{eqnarray}
where $\rho$ is the density, $\bfv$ is the velocity, $g$ is the 
gravitational acceleration, $p$ is the pressure, $T$ is the temperature, 
$s$ is the entropy per unit mass, $\epsilon$ is the nuclear energy 
generation rate,  
\begin{equation}\label{radiativefluxeq}
\bffr = -\frac{ac}{3 \kappa \rho} \nabla T^{4}
\end{equation}
is the radiative flux, $a$ is the radiation constant, and $\kappa$ is
the radiative opacity.  $g$ is presumed constant since the vertical
thickness of the convective layer is small relative to the stellar
radius.  Equations (\ref{continuityeq}-\ref{entropyeq}) govern the
aggregate behavior of the burning layer during a type I X-ray bursts.
Specifically, they describe both the oscillatory and convective
motions of a matter element.  To proceed further I need to derive both
the equations that govern only the long wavelength, oscillatory
motions and the equations that govern only the short wavelength,
convective motions.  I do so in the following subsection.

\subsection{Oscillatory and Convective Equations}

Motivated by the work of \citet{U67} and \citet{GSNB75}, I decompose
the physical quantities into their oscillatory and convective parts:
\begin{eqnarray}\label{variables}
\rho &=& \rhobar + \delrho, \nonumber\\ p &=& \pbar +
\delp,\nonumber\\ T &=& \Tbar + \delT,\nonumber\\ s &=& \sbar +
\dels,\nonumber\\ \bfv &=& \vbar + \delv,
\end{eqnarray}
where $\rhobar$, $\pbar$, $\Tbar$, $\sbar$, and $\vbar$ are values
averaged over a horizontal area with dimensions much larger than the
mixing length but much smaller than the oscillation wavelength, and
$\delrho$, $\delp$, $\delT$, $\dels$, and $\delv$ are the local
quantities that describe the convective fluctuations.  The convective
quantities are much smaller than their respective oscillatory
quantities except for the velocity, in which case $|\delv| \gg |\vbar|$.
In fact, $\vbar = 0$ in the unperturbed configuration.  Note that 
$\overline{\Delta x} = 0$ for all quantities $x$.  
Furthermore, I
define the Lagrangian derivative following the oscillatory motion
\begin{equation}\label{lagrangianderivative}
\ddt{} \equiv \ppt{} + \vbar \cdot \nabla.
\end{equation}
For each of the three fundamental equations
(\ref{continuityeq}-\ref{entropyeq}), I first derive the oscillatory
equations by taking the horizontal average of the corresponding
fundamental equation.  I then derive each convective equation by
subtracting the corresponding oscillatory equation from the
fundamental equation.

\subsubsection{Continuity Equation}

Implementing equations (\ref{variables}) and (\ref{lagrangianderivative}), 
I write the continuity equation (\ref{continuityeq}) as
\begin{equation}
\ddt{(\rhobar + \delrho)} + (\rhobar + \delrho)\nabla \cdot \vbar + 
\nabla \cdot (\overline{\rho \delv}) = 0.
\end{equation}
Setting $\delrho = 0$ by assumption (I) and noting that 
$\overline{\rho \delv} = 0$ \citep[since convection involves no 
bulk motion, e.g.][]{G96}, the space-averaged continuity 
equation becomes
\begin{equation}\label{continuityeqosc}
\ddt{\rhobar} + \rhobar \nabla \cdot \vbar = 0.
\end{equation}
Subtracting equation (\ref{continuityeqosc}) from (\ref{continuityeq})
and again implementing assumption (I) ($\delrho = 0$, $\rhobar =
\mathrm{constant}$) gives the continuity equation describing the
convective motions,
\begin{equation}\label{continuityeqcon}
\nabla \cdot \delv = 0.
\end{equation}
 
\subsubsection{Momentum Equation}

Taking the horizontal average of equation (\ref{momentumeq}) gives
\begin{equation}\label{momentumeqosc}
\ddt{\vbar} + \overline{(\delv \cdot \nabla) \vbar} + \overline{(\delv \cdot \nabla) \delv} + 2\Omega \times \vbar = 
- \frac{1}{\rhobar} \nabla \pbar - g\zhat.
\end{equation}
Subtracting equation (\ref{momentumeqosc}) from (\ref{momentumeq}) 
gives the momentum equation for the convective motions,
\begin{eqnarray}
\ddt{\delv} + (\delv \cdot \nabla)\delv - \overline{(\delv \cdot
\nabla) \delv} + (\delv \cdot \nabla)\vbar - \overline{(\delv \cdot
\nabla) \vbar} 
\nonumber \\ \nonumber \\
+ 2\Omega \times \delv 
= -\frac{1}{\rhobar} \nabla
(\delp) + \frac{\delrho}{\rhobar^{2}} \nabla \pbar.
\end{eqnarray}
To be consistent with the mixing length theory, I set 
\begin{eqnarray}\label{mlteq}
(\delv \cdot \nabla)\delv -  \overline{(\delv \cdot \nabla) \delv }
&\equiv& \Lambda \frac{\delv}{\tauc}, \nonumber \\
(\delv \cdot \nabla)\vbar -  \overline{(\delv \cdot \nabla) \vbar }
&\equiv& \Lambda \frac{\vbar}{\tauc},
\end{eqnarray}
where 
\begin{equation}\label{tauceq}
\tauc = \frac{l}{|V_z|}
\end{equation}
is the lifetime of a convective element, $V_z = \delv \cdot \zhat$,
and $\Lambda$ is a number of order unity \citep{B58,U67,GDGGS05}.  The
momentum equation for convection becomes
\begin{equation}\label{momentumeqcon}
\ddt{\delv} + 2\Omega \times \delv = -\frac{1}{\rhobar} \nabla (\delp)
+ \frac{\delrho}{\rhobar^{2}} \nabla \pbar - \Lambda
\frac{\vbar}{\tauc} - \Lambda \frac{\delv}{\tauc}.
\end{equation}

\subsubsection{Entropy Equation}

Taking the horizontal average of equation (\ref{entropyeq}) gives
\begin{equation}\label{entropyeqosc0}
\rhobar \Tbar \ddt{\sbar} = \rhobar \bar{\epsilon} - \nabla \cdot \bffrbar 
- \overline{\rho T \delv \cdot \nabla s}.
\end{equation}
I need to express the last term in a more useful form.  Using the
thermodynamic identity $Tds = dH - dp/\rho$, where $H$ is the enthalpy
per unit mass, I find
\begin{equation}
\overline{\rho T \delv \cdot \nabla s} = \overline{\rho \delv \cdot
\nabla H} - \overline{\delv \cdot \nabla p}.
\end{equation}
$\overline{\delv\cdot \nabla p}$ is the turbulent
kinetic energy dissipation rate and is negligible by assumption (IV),
so I omit it from the subsequent analysis.  Assumption (II) implies
that $\nabla H = \nabla(\bar{H} + \Delta H) \approx \nabla \Delta H$.
Using this and equation (\ref{continuityeqcon}), I find
\begin{equation}
\overline{\rho T \delv \cdot \nabla s} = \nabla \cdot \overline{\rho
\delv \Delta H}.
\end{equation}
In general, $\Delta H = T \dels + \delp/\rho$ by the thermodynamic
identity, but since
$\delp = 0$ by assumption (I), $\Delta H = T \dels$, so the 
horizontally-averaged entropy equation becomes
\begin{eqnarray}
\rhobar \Tbar \ddt{\sbar} &=& \rhobar \bar{\epsilon} - \nabla \cdot (\bffrbar 
+ \bffcbar), \label{entropyeqosc} \\ 
\bffcbar &\equiv& \overline{\rho \delv T \dels}, \label{convectivefluxeq}
\end{eqnarray}
where $\bffcbar$ is the horizontally-averaged convective flux.  I use
equations (\ref{entropyeqosc}) and (\ref{convectivefluxeq}) in \S
\ref{perturbeqns}.  Subtracting equation (\ref{entropyeqosc0}) from
(\ref{entropyeq})  convective entropy equation
\begin{eqnarray}
(\delrho \Tbar + \rhobar{\delT}) \ddt{\sbar} 
+ \rhobar \Tbar \ddt{\dels} + \rho T \delv \cdot \nabla s - 
\overline{\rho T \delv \cdot \nabla s} 
\nonumber \\ \nonumber \\
= \rho \epsilon - \rhobar 
\bar{\epsilon} - \nabla \cdot(\bffr-\bffrbar).
\end{eqnarray}
To simplify the equation and keep it consistent with the mixing length 
theory, I set, in analogy with equations 
\citep[\ref{mlteq}; see][]{GDGGS05},
\begin{equation}
\delv \cdot \Delta (\rho T \nabla s) - \overline{\delv \cdot (\rho T
\nabla s)} = \rhobar \Tbar\frac{\dels}{\tauc}.
\end{equation}
Setting $\Delta (\rho\epsilon) = 0$ and $\Delta \bffr = 0$ by assumption 
(III) gives the convective entropy equation
\begin{equation}\label{entropyeqcon}
\left(\frac{\delrho}{\rhobar}+\frac{\delT}{\Tbar}\right)\ddt{\sbar} + 
\ddt{\dels} + \delv \cdot \nabla \sbar = -\frac{\dels}{\tauc}.
\end{equation}

\section{Perturbation of the Height-Integrated Equations}\label{perturbeqns}

For simplicity, I construct a one-zone model for the vertical
structure of the accreted layer \citep[e.g.][]{FHM81,B98,NC07}.  
I assume that the horizontally-averaged quantities describing
 the accreted matter are 
constant throughout the layer.
By assumption (V), I 
write the height of the accreted layer as 
\begin{equation}\label{h}
\barh = \frac{\pbar}{\rhobar g}.
\end{equation}
The equation of state is  
\begin{eqnarray}
p &=& \frac{\rho \kB T}{\mu m_p} + \frac{aT^{4}}{3},\label{eos}\\
s &=& \frac{\kB}{\mu m_p} \left[ \ln\left(\frac{T^{3/2}}{\rho}
\right) + \frac{4}{\beta}-4 \right] + \mathrm{constant},
\end{eqnarray}
where $\kB$ is Boltzmann's constant, $\mu$ is the mean molecular weight, 
$m_p$ is the proton mass, and 
$\beta = \rho \kB T/\mu m_p p$ is the ratio of gas to total pressure.  
Perturbing the equation of state gives
\begin{equation}\label{dtoverteq}
\dtovert = \frac{1}{\chi_T} \dpoverp 
- \frac{1}{\upsilon_T} \drhooverrho,
\label{Teq}
\end{equation}
where $\chi_T = 4-3\beta$ and 
\begin{equation}
\upsilon_T \equiv - \left (\frac{\partial \ln \rho}{\partial \ln T}\right)_P = \frac{4-3\beta}{\beta}
\end{equation}
is minus the coefficient of thermal expansion.
From equations (\ref{radiativefluxeq}), (\ref{entropyeqosc}), and
(\ref{convectivefluxeq}), the height-integrated entropy equation is
\begin{eqnarray}
\pbar \Tbar \ddt{\sbar} &=& \pbar \bar{\epsilon} - g(\frbar + 
\fcbar)\label{hintentropyeq},\\
\fr &=& \frac{acgT^{4}}{3 \kappa p},\label{freq}\\
\fc &=& \rho T V_z \Delta s\label{fceq}.
\end{eqnarray}
The goal of this work is to perturb equation
(\ref{hintentropyeq}) and determine the stability of the
burning layer to nonradial oscillations.  However, the 
expression for the perturbed convective flux $\delta \fcbar$ 
includes perturbations of convective quantities.  To proceed, I
first need to express the perturbed convective quantities $\delta
V_z$ and $\delta \dels$ in terms of the perturbed oscillatory
quantities $\delta \rhobar$, $\delta \Tbar$, and $\delta \pbar$.  The
following subsection is devoted to this.

\subsection{Perturbed Convective Equations}

I conduct a standard WKB perturbation analysis on the convective
equations.  I assume that all dynamical quantities vary as $\exp(-i
\omega t + i \bfk \cdot {\bf x})$, where $\bfk = k_x \xhat + k_y \yhat
+ k_z \zhat$ and $k \barh \ll 1$.  Since the lifetime of a convective
element $\tauc$ is assumed to be much smaller than the mode period $2
\pi/\omega$, I presume that all unperturbed convective quantities are
in quasi-steady state.

Deriving the perturbed continuity equation for convection is simple.  
From equation (\ref{continuityeqcon}), the continuity equation 
for convective motions is 
\begin{equation}
\bfk \cdot \delv = 0,\label{kdotv}
\end{equation}
so the perturbed continuity equation for convection is 
\begin{equation}
\bfk \cdot \delta \delv = 0\label{kdotdeltav}.
\end{equation}

Invoking the WKB approximation, integrating over $z$, and simplifying,  
the momentum equation for convection (\ref{momentumeqcon}) becomes
\begin{equation}\label{intmomeqcon}
\ddt{\delv} + 2 \Omega \times \delv = -i \bfk \frac{\delp}{\rhobar} +
\frac{\delrho}{\rhobar} g \zhat - \Lambda
\frac{\vbar}{\tauc} - \Lambda
\frac{\delv}{\tauc},
\end{equation}
where I have assumed that $\nabla \pbar \approx (\partial
\pbar/\partial z) \zhat$ in the unperturbed state.  Perturbing
equation (\ref{intmomeqcon}), setting both $d/dt \rightarrow 0$ and
$\Omega \tauc \rightarrow 0$ by assumption (III), and ignoring the
term $\Lambda \vbar/\tauc$ because $|\delv| \gg |\vbar|$ gives the
perturbed convective momentum equation
\begin{equation}\label{perturbedmomeqcon}
i \bfk \delta \left(\frac{\delp}{\rhobar g}\right)  + \frac{\Lambda}{g \tauc}
\left(\delta \delv - \delv \frac{\delta \tauc}{\tauc} \right) = \delta \left(
\frac{\delrho}{\rhobar} \right) \zhat.
\end{equation}
Equation (\ref{perturbedmomeqcon}) contains the perturbed 
convective quantities $\delta \delp$, $\delta \delrho$, $\delta \delv$, 
and $\delta \tauc$, whereas the desired perturbed quantities are 
$\delta \dels$ and $\delta V_z$.  To remedy this, I proceed as follows.  
Taking the dot product of equation (\ref{perturbedmomeqcon}) and 
$\bfk$ and using equations (\ref{kdotv}) and (\ref{kdotdeltav}), I find
\begin{equation}\label{delpeq}
\delta \left( \frac{\delp}{\rhobar g}\right) = \delta
\left(\frac{\delrho}{\rhobar} \right) \frac{k_z}{i k^{2}}.
\end{equation}
Taking the dot product of equation (\ref{perturbedmomeqcon}) and 
$\zhat$ and using equation (\ref{delpeq}) then gives
\begin{equation}\label{vreq}
\frac{\Lambda V_z}{g \tauc} \left(\frac{\delta V_z}{V_z} -
\frac{\delta \tauc}{\tauc}\right) = \delta
\left(\frac{\delrho}{\rhobar} \right )
\left(1-\frac{k_z^{2}}{k^{2}}\right).
\end{equation}
Conducting a similar procedure on the unperturbed, steady-state
convective momentum equation, I find
\begin{equation}
 \frac{\Lambda V_z}{g \tauc} = \left(\frac{\delrho}{\rhobar} \right ) 
\left(1-\frac{k_z^{2}}{k^{2}}\right).
\end{equation}
It follows that 
\begin{equation}
\frac{\delta V_z}{V_z} = \frac{\delta \tauc}{\tauc} + \frac{\delta 
(\delrho/\rhobar)}{\delrho/\rhobar}.
\end{equation}
To proceed further, I need expressions for the two terms on the 
right hand side of the above equation.  From equation (\ref{tauceq}), 
\begin{equation}\label{deltatauceq0}
\frac{\delta \tauc}{\tauc} = \frac{\delta l}{l} - \frac{\delta V_z}{V_z}.
\end{equation}
I make the standard assumption that the mixing length $l$ is
proportional to the pressure scale height $\barh$
\citep[e.g.,][]{CG68,HKT04}.  The lifetime of a convective element is
negligible compared to the oscillation period (i.e.~$\omega \tauc \ll
1$) by assumption (III).  Therefore, I assume the mixing length 
instantaneously adjusts to changes in the scale height.  
It follows that
\begin{equation}\label{deltaleq}
\frac{\delta l}{l} = \frac{\delta \barh}{\barh} = \dpoverp -
\drhooverrho,
\end{equation}
where the last equality follows from equation (\ref{h}).  Note that
equation (\ref{deltaleq}) would not apply if, for example, $\omega \tauc \gg
1$; in that case, $\delta l/l = 0$.  Equations (\ref{deltatauceq0})
and (\ref{deltaleq}) then give
\begin{equation}\label{deltatauceq}
\frac{\delta \tauc}{\tauc} = \dpoverp - \drhooverrho - \frac{\delta V_z}{V_z}.
\end{equation}
Writing the convective entropy as $T\dels = c_p [\delT -(\partial
T/\partial p)_s \delp]$, setting $\delp = 0$ by assumption (I), and
noting that $\delT /\Tbar = - (1/\upsilon_T) \delrho/\rhobar$ from the
equation of state, I get
\begin{equation}
\frac{\delrho}{\rhobar} = - \upsilon_T \frac{\dels}{c_p},
\end{equation}
where 
\begin{equation}
c_p = \frac{5\kB}{2\mu m_p}\left(\frac{32-24\beta-3\beta^{2}}{5\beta^{2}}
\right)
\end{equation}
is the specific heat at constant pressure.  Thus, I find that
\begin{equation}\label{deltavzeq}
\frac{\delta V_z}{V_z} = \frac{1}{2} \left(\dpoverp - 
\drhooverrho+ \frac{\delta \dels}{\dels} + \frac{\delta 
\upsilon_T}{\upsilon_{T}} - \frac{\delta c_p}{c_{p}} \right).
\end{equation}
All I need now is an expression for $\delta \dels$ in terms of
the oscillatory variables.

Integrating the convective entropy equation (\ref{entropyeqcon}) over
$z$  gives
\begin{equation} 
\left(\frac{\delrho}{\rhobar}+\frac{\delT}{\Tbar}\right)\ddt{\sbar} + 
\ddt{\dels} + \frac{V_z \sbar \rhobar g}{\pbar} = -\frac{\dels}{\tauc}.
\end{equation} 
Perturbing this equation and setting $\omega \tauc \rightarrow 0$ 
by assumption (III), I find
\begin{equation}
\frac{\delta V_z}{V_z} + \dsovers + \drhooverrho - \dpoverp = \frac{\delta
\dels}{\dels} - \frac{\delta \tauc}{\tauc}.
\end{equation}
Using equations (\ref{deltatauceq}) and (\ref{deltavzeq}), the above
relation simplifies to
\begin{equation}\label{deltadelseq}
\frac{\delta \dels}{\dels} = \frac{\delta \sbar}{\sbar}.
\end{equation}

\subsection{Perturbed Entropy Equation}

Perturbing the left side of the 
horizontally-averaged entropy equation (\ref{entropyeqosc}) 
and using equation (\ref{dtoverteq}) to eliminate $\delta \Tbar/\Tbar$, 
I find
\begin{equation}\label{deltaTdsdt}
\delta \left(\pbar \Tbar \ddt{\sbar}\right) = -i \omega 
\frac{\pbar g \barh}{(\Gamma_3 - 1)}\left(\dpoverp - \Gamma_1 \drhooverrho\right),
\end{equation}
where $\Gamma_1 = (\partial \ln p/\partial \ln \rho)_s$ and 
$\Gamma_3 - 1 = (\partial \ln T/\partial \ln \rho)_s$ are the 
usual adiabatic exponents \citep[e.g.][]{HKT04}.  The nuclear energy 
generation rate $\epsilon$ is a function of both density 
and temperature.  For small perturbations 
about the initial configuration, I write 
\begin{equation}
\delta \bar{\epsilon} = \Omega_h gh_0 \left (\nu \dtovert + 
\eta \drhooverrho \right ), \, \nu \equiv \left (\frac{\partial \ln \epsilon}{\partial \ln T}\right)_\rho, \, \eta \equiv \left (\frac{\partial \ln \epsilon}{\partial \ln \rho}\right)_T,
\end{equation}
where $\Omega_h^{-1}$ is the characteristic heating time of the layer
via nuclear burning.  Using equation (\ref{dtoverteq}) to replace
$\delta \Tbar/\Tbar$ in favor of $\delta \pbar/\pbar$, the perturbed
heating term becomes
\begin{equation}\label{deltaepsilon}
\delta \epsilon = \Omega_h g \barh \left[\frac{\nu}{\chi_T} \dpoverp +
\left(\eta-\frac{\nu}{\upsilon_T}\right)\drhooverrho \right].
\end{equation}
The layer cools via both radiative diffusion and convection.  The
radiative flux $F_R$ depends on the radiative opacity $\kappa$, which
is a function of both density and temperature.  For small
perturbations about the initial configuration, I write
\begin{equation}\label{kappaeq}
\frac{\delta \kappa}{\kappa} = \zeta \dtovert + 
\xi \drhooverrho, \quad \zeta \equiv \left (\frac{\partial \ln \kappa}{\partial \ln T}\right)_\rho, \quad \xi \equiv \left (\frac{\partial \ln \kappa}{\partial \ln \rho}\right)_T.
\end{equation}
Note that equations (\ref{deltaTdsdt}-\ref{kappaeq}) are simply
generalizations of their analogous equations of \citet{NC07}.  The
magnitude of the unperturbed convective flux $\fcbar$ depends on the
convective quantities $V_z$ and $\Delta s$.  I cannot determine their
values in terms of the oscillatory quantities $\rhobar$, $\Tbar$, and
$\pbar$ from first principles, so the magnitude of $\fcbar$ is
uncertain.  I parameterize this uncertainty by defining the
dimensionless quantity
\begin{equation}\label{fconveq}
\fconv \equiv \frac{\fcbar}{\frbar+\fcbar},
\end{equation}
the ratio of the unperturbed convective flux to the unperturbed 
total flux.  Using equations (\ref{freq}), (\ref{fceq}), 
(\ref{kappaeq}), and (\ref{fconveq}), I find
\begin{eqnarray}
\delta (\frbar + \fcbar) &=& \Omega_c \pbar \barh 
\left \{(1-\fconv)\left[(4-\zeta)\dtovert - \dpoverp - \xi
\drhooverrho \right] 
\right. \nonumber \\ \nonumber \\ && \left. 
+ \fconv \left(\drhooverrho + \dtovert + \frac{\delta V_z}{V_z} + \frac{\delta \dels}{\dels}
\right) \right \},
\end{eqnarray}
where $\Omega_c^{-1}$ is the characteristic cooling time of the layer.
Finally, using equations (\ref{deltavzeq}) and (\ref{deltadelseq}) to 
eliminate the perturbed convective terms and using 
equation (\ref{dtoverteq}) to eliminate $\delta \Tbar/\Tbar$, the 
perturbed cooling term becomes
\begin{eqnarray}\label{deltaflux}
\delta (\frbar + \fcbar) &=& \Omega_c \pbar
\barh \left\{ \left[\frac{4-\zeta}{\chi_T} - 1 + \fconv
\left(\frac{3}{2} - \frac{3-\zeta}{\chi_T}\right)\right]\dpoverp \right. 
\nonumber \\ \nonumber \\ &&\left. -
\left[\frac{4-\zeta}{\upsilon_T} + \xi - \fconv \left(\frac{1}{2} +
\frac{3-\zeta}{\upsilon_T}+ \xi \right) \right]\drhooverrho 
\right. \nonumber\\ \nonumber \\ &&\left.
+ \frac{\fconv}{2}\left( 3\frac{\delta \sbar}{\sbar} + \frac{\delta
\upsilon_T}{\upsilon_T} - \frac{\delta c_P}{c_P}\right) \right \}.
\nonumber \\ \nonumber \\
\end{eqnarray}
Substituting equations (\ref{deltaTdsdt}), (\ref{deltaepsilon}), and
(\ref{deltaflux}) into the perturbed entropy equation
\begin{equation}
\delta \left(\pbar \Tbar \ddt{\sbar}\right) = \delta (\pbar \epsilon)- g 
\delta(\frbar + \fcbar)
\end{equation}
and simplifying gives and expression for $\delta \pbar$ in terms of 
$\delta \rhobar$,
\begin{equation}
\dpoverp = A \drhooverrho,
\end{equation}
where $A$ is a dimensionless complex quantity.  For purely adiabatic
perturbations, $A = \Gamma_1$.  In the limit where the convective
layer is gas pressure dominated ($\beta \rightarrow 1$)
\begin{equation}\label{Abeta1}
A = \frac{5}{3}\left(\frac{1 + (2i/5)(\Omega_c/\omega)[4-\zeta+\xi-
\fconv(7/2+\xi-\zeta)-(\Omega_h/\Omega_c)(\nu-\eta)]}{1 + 
(2i/3)(\Omega_c/\omega)
[3-\zeta-\fconv(3/2-\zeta)-(\Omega_h/\Omega_c)(1+\nu)]}\right),
\end{equation}
and where the layer is radiation pressure dominated ($\beta \rightarrow 0$)
\begin{equation}\label{Abeta0}
A = \frac{4}{3}\left(\frac{1 + (i/4)(\Omega_c/\omega)[\xi+
\fconv(1/2-\xi)+(\Omega_h/\Omega_c)\eta]}{1 + (i/3)(\Omega_c/\omega)
[-\zeta/4+\fconv(3/2+\zeta/4)-(\Omega_h/\Omega_c)(1+\nu/4)]}\right).
\end{equation}
I work exclusively in these two limits hereafter for simplicity.

\section{Pulsational Stability}\label{pulsationalstab}

I invoke the quasi-adiabatic approximation to determine the linear
stability of the layer.  That is, I presume the fractional
entropy change during one oscillation period is small.  An equivalent
statement is that the heating and cooling rates are small relative to
the oscillation frequency, i.e. $\Omega_h/\omega \ll 1$ and
$\Omega_c/\omega \ll 1$.  The typical heating and cooling times during
a type I X-ray burst are of order one second, so $\Omega_h$ and
$\Omega_c \sim 1\, \mathrm{s}^{-1}$, whereas the modes have angular
frequencies $\sim 100\, \mathrm{radians}\,\mathrm{s}^{-1}$
\citep[e.g.,][]{H04}, so the quasi-adiabatic approximation is valid.

I use the following criterion to determine the pulsational stability
of the convective layer: it is unstable if the
work done on the layer during one oscillation period is positive, and
it is stable if the work done on the layer during one oscillation
period is negative \citep{C80,UOASS89}.  The pressure and density
perturbations both vary in time as $\exp(-i\omega t)$, but they are in
general out of phase because $A$ is complex.  If $\mathrm{Im}\,(A) >
0$, $\delta \pbar$ lags behind $\delta \rhobar$ and the area
traced in the $P$--$V$ diagram during one oscillation period is
positive; the $PdV$ work done on the layer is positive and the layer
is unstable.  Conversely, if $\mathrm{Im}\,(A) < 0$,
$\delta \pbar$ leads $\delta \rhobar$ and the area traced in the
$P$--$V$ diagram is negative; the $PdV$ work done on
the layer is negative and the layer is stable.  For purely adiabatic
perturbations, $A$ is real and $\delta \pbar$ and $\delta \rhobar$
are in phase.  The time evolution of $\delta \pbar$ and $\delta \rhobar$ 
traces a curve of zero area in the $P$--$V$ diagram, so the $PdV$ work done in one cycle is zero; the
modes neither grow nor decay.  Thus the sign of $\mathrm{Im}\,(A)$
determines the stability of the layer.

At the high temperatures achieved during a burst, the opacity
scales according to the approximate formulae given by \citet{Pre83},
which give $-0.1 \lesssim \xi \lesssim 0$, $-0.5 \lesssim \zeta
\lesssim 0$.  I follow \citet{NC07} and set $\xi=0$ and $\zeta=-0.25$
for simplicity.  Solving for the imaginary part of $A$ in equations
(\ref{Abeta1}) and (\ref{Abeta0}), I find the instability
criterion for nonradial oscillations to be
\begin{equation}\label{stabbeta1}
\frac{\Omega_h}{\Omega_c} > \frac{7 + 5 \fconv}{10 + 6\eta + 4\nu}
\end{equation}
when the layer is gas pressure dominated ($\beta \rightarrow 1$), and
\begin{equation}\label{stabbeta0}
\frac{\Omega_h}{\Omega_c} > \frac{1 + 17 \fconv}{16 + 12\eta + 4\nu}
\end{equation}
when the layer is radiation pressure dominated ($\beta \rightarrow
0$).  Note that equations (\ref{stabbeta1}) and (\ref{stabbeta0}) reduce to
equations (76) and (77) of \citet{NC07} when $\fconv = 0$, as they
should.  If the heating rate due to thermonuclear burning during a
burst is sufficiently large relative to the cooling rate, some of the
nuclear energy converts to mechanical energy and drives the surface
modes.  It is clear from equations
(\ref{stabbeta1}) and (\ref{stabbeta0}) that convection dampens oscillations
in both cases.  For $\beta = 1$, the effect is fairly modest; the
minimum heating rate needed for instability when cooling is
convection-dominated is less than a factor of $2$ greater than that
needed when cooling is radiation-dominated.  However, the temperature
of the layer during a burst is often $\gtrsim 10^{9}\, \mathrm{K}$,
especially when a sizable convective layer develops
\citep{J77,Wetal04}, so the case $\beta = 0$ is more relevant
for this work.  For $\beta = 0$, the effect is huge; the minimum
heating rate needed for instability when cooling is
convection-dominated is $18$ times greater than that needed when
cooling is radiation-dominated.  Therefore, oscillations during type I
X-ray bursts are unlikely when convection dominates the cooling.

\citet{L04} draws the opposite conclusion: nuclear burning drives
oscillations in a convective layer.  Focusing primarily on the mode
structure instead of the stability criterion, Lee simplifies his
calculation by assuming $\delta (\nabla \cdot \bffcbar) = 0 $, which
implies that the convection zone is ``frozen'' during the
perturbation.  This assumption is inapplicable to type I X-ray bursts
because $\omega \tauc \ll 1$, i.e. the convection zone has sufficient
time to respond to oscillatory motions.

\section{Discussion}\label{discuss}

In this investigation, I have found that efficient convection dampens
nonradial oscillations during type I X-ray bursts.  This may explain
the nondetection of oscillations near the peak of some bursts.  The
basic physics of this effect is simple.  Consider the lateral
compression of a column of matter during a burst.  When the matter is
compressed, the temperature and density rise and generally increase
the nuclear energy generation rate; the cooling rate may either
increase or decrease.  If the marginal increase in the heating rate is
sufficiently large relative to the cooling rate at maximum
compression, the pressure will continue to rise and thereby drive the
oscillations \citep[see also \S 5.2 of ][]{NC07}.  Energy transport by
convection is more effective than that by radiative diffusion when the
layer is compressed, so the cooling rate is larger at maximum
compression if the layer is convective.  In this case, the pressure is
more likely to decrease at maximum compression and hence dampen the
oscillations.

The convection zone lasts only a few seconds after ignition and
disappears roughly when the burst luminosity reaches its peak.  The
$\epsilon$-mechanism may drive oscillations at this time.  The
$r$-mode growth timescale is a few seconds when the burning layer is
radiative \citep{NC07}, so oscillations may be detected in the burst
tail even if convection dampens them near the peak.

It is unclear if convection alone explains the paucity of burst
oscillation detections during PRE.  Strong convection is guaranteed
during a PRE burst since the nuclear burning timescale is much less
than the radiative cooling timescale, but the heating rate is probably
large enough relative to the convective cooling rate to satisfy the
instability criterion.  It may be that, although convection cannot
damp the oscillations entirely, it lowers the growth rate enough to
keep the oscillation amplitudes below the detectability threshold.
Alternatively, perhaps the surface modes are unstable during PRE, but
the bursting layer's vertical extent is sufficiently large relative to
the mode wavelength to smear out the nonaxisymmetry \citep{CB00}.

I have presented a simplified, analytical, one-zone model of the
convective burning layer as a first attempt to study the effect
convection has on the driving of burst oscillations.  There are
several important issues I am unable to address.  First, convection
alters both the energy transport mechanism and the temperature
gradient.  I have investigated only the former in this work.  Second,
the model assumes the convection zone's extent is constant in time.
During an actual burst, the convection zone first grows to lower
pressures and then recedes to the ignition region
\citep[e.g.,][]{J78,Wetal04,WBS06,FST08}.  Presumably, the convection
zone in a given column of matter will expand when the column is
compressed and thereby further dampen the oscillations, but this is
merely a speculation.  Third, I cannot determine the actual values of
the critical parameters $\Omega_h/\Omega_c$, $\fconv$, $\nu$, and
$\eta$ of the stability criterion.  One can determine them only with
detailed, multi-zone calculations such as those of \citet{Wetal04} and
\citet{FST08}.

\acknowledgments 

I thank Ramesh Narayan for his advice and encouragement, Duncan
Galloway and Mike Muno for answering my questions about the latest
observations of burst oscillations, Deepto Chakrabarty and Tony Piro
for helpful discussions, Nevin Weinberg for reviewing the
manuscript, and the referee for questions and comments that helped me 
improve the text.  NASA grant NNG04GL38G supported this work.

\bibliographystyle{apj}

\end{document}